# From Periodicity to Chaos: Stabilization, Bifurcation, and Synchronization of the Van der Pol Oscillator

Faheem Ahmed Chughtai
*Department of Mechanical Engineering, College of Electrical & Mechanical Engineering, National University of Sciences and Technology, Islamabad, 44000, Pakistan*
*fahmed.me22ceme@student.nust.edu.pk*

Imran Akhtar
*Department of Mechanical Engineering, College of Electrical & Mechanical Engineering, National University of Sciences and Technology, Islamabad, 44000, Pakistan*
*imran.akhtar@ceme.nust.edu.pk*

Imran Shafi
*Department of Mechanical Engineering, College of Electrical & Mechanical Engineering, National University of Sciences and Technology, Islamabad, 44000, Pakistan*
*dacad.ceme@ceme.nust.edu.pk*

***Abstract*— This work investigates the dynamics of the van der Pol oscillator, a well-known dynamical model used to represent many naturally occurring phenomena, under both unforced and externally forced conditions, with a focus on its stability characteristics and bifurcation behavior. An approximate analytical solution is derived using the Method of Multiple Scales (MMS) and validated against numerical simulations conducted via the ode45 solver. The evolution of the system's limit cycle is examined as the strength of the nonlinear damping term increased. A range of bifurcation scenarios is explored by systematically varying relevant control parameters. The system's response to external forcing at different damping levels is analyzed to uncover transitions toward chaotic behavior and subsequent phase locking (entrainment). Finally, the relevance of the van der Pol oscillator as a model for naturally occurring rhythmic or periodic processes is discussed, highlighting its applicability in representing biological and physical systems.**

## I. Introduction

The Van der Pol equation (VPE) is a nonlinear second-order differential equation that was originally formulated by the Dutch physicist Balthasar Van der Pol in the early 20th century. Since its introduction, the model has been widely adopted to describe self-sustained oscillatory behavior in a variety of physical and biological systems. Its relevance spans across multiple disciplines, including electrical engineering, acoustics, biomedical sciences, and nonlinear physics. Among its notable biomedical applications is the approximation of cardiac dynamics, where the VPE—especially under conditions of strong nonlinear damping—has been employed to capture features of the human heartbeat. This analogy has been further extended to represent pulsatile blood flow dynamics within the aorta [1]. More recently, the Van der Pol framework has been combined with data-driven techniques, including machine learning and evolutionary optimization algorithms, to identify parameters capable of reproducing the morphology of actual electrocardiographic (ECG) signals [2].

The analysis of the Van der Pol equation (VPE) using perturbation methods, particularly the Method of Multiple Scales (MMS), offers an effective strategy for resolving the system dynamics across different temporal scales. This technique enables the derivation of approximate analytical solutions and facilitates a parametric investigation of the nonlinear system behavior [3][4]. In the context of autonomous dynamics, Hafeez et al. demonstrated the presence of a unique limit cycle by examining fixed points and their stability characteristics [5]. In the domain of nonlinear aeroelastic phenomena such as vortex-induced vibrations (VIV), galloping, and flutter, Guangzhong Gao and colleagues illustrated that both Van der Pol and Rayleigh-type formulations serve as generalized models for capturing the amplitude behavior of self-excited oscillations [6]. The system's bifurcation landscape—encompassing stable, unstable, and Hopf bifurcations—has been extensively studied to characterize the emergence of limit cycles as the nonlinear damping parameter μ varies [7][8][9]. Under strong nonlinear forcing, the system often transitions into chaotic regimes, which introduces complexities in designing effective control strategies [10]. Interestingly, the inherent chaotic dynamics of the Van der Pol oscillator have also been applied in secure communication schemes, including the development of image encryption algorithms whose validity has been confirmed through simulation-based evaluations [11].

This study investigates the Van der Pol equation (VPE) with a focus on its sensitivity to changes in system parameters, aiming to assess its effectiveness as a nonlinear oscillator and limit cycle generator. Section II(A) outlines the analytical methodology employed to solve the VPE and derives solutions across varying orders of approximation. The analytical expressions serve to support the parametric analysis by highlighting the influence of key variables on the system's response. Section II(B) presents a detailed fixed-point and stability analysis, supported by numerical simulations to examine the emergence of both stable and unstable limit cycles and their relationship with equilibrium point behavior. Section III delves into bifurcation characteristics of the nonlinear system, utilizing Floquet multiplier theory to track qualitative changes in the system's dynamics, along with spectral analysis to evaluate system behavior across bifurcation thresholds. Section IV extends the investigation to the forced form of the VPE, examining transitions to various bifurcation types under co-dimensional parameter variations. It also explores synchronization with external forcing and identification of frequency-locking regions using Arnold tongue representation of VPE, and the onset of chaos using Lyapunov exponent theory. Section V describes the major application areas of the Van der Pol dynamical system to naturally occurring phenomena and its application in modelling the natural systems. Section VI and VII highlights the comparative analysis and certain limitations of the proposed framework and finally, Section VIII summarizes the key findings and discusses the broader relevance of the VPE as a representative model for complex oscillatory dynamics.

## II. Analytical Solution

In this work, we investigated the influence of nonlinearity on the self-excited Van der Pol oscillator, as well as its response to external forcing. To generate analytical correlation of the system dynamics against full numerical solution, we approximated the solution using MMS, which allows a systematic exploration of the system's dynamics under weak nonlinear conditions. The general form of the VPE considered in this study is expressed as:

$$\ddot{x} + \omega_0^2 x = \epsilon(1 - x^2)\dot{x} + F(\cos\omega t) \tag{1}$$

An examination of the right-hand side of the Van der Pol equation—excluding the external forcing term—reveals a key nonlinear damping behavior. For small oscillation amplitudes, the variable $x$ remains small, resulting in a negative damping effect. This implies that energy is being injected into the system, which can drive the response toward instability. Conversely, when the amplitude becomes large, the damping becomes positive, leading to energy dissipation. This dual behavior highlights a distinctive feature of the VPE: its nonlinear damping mechanism supplies energy at low amplitudes and dissipates it at higher amplitudes. This self-regulating energy exchange enables the system to sustain a stable limit cycle over time.

### *A. Application using Method of Multiple Scales (MMS)*

The Method of Multiple Scales (MMS), as introduced by Nafeh, is a perturbation technique particularly effective for analyzing weakly nonlinear systems like the Van der Pol equation. In such systems, the solution ***x(t)*** typically exhibits dependencies on both fast and slow timescales—namely, the natural time variable $t$ and the nonlinear perturbation parameter $\epsilon$. Since $x$ is influenced by combinations like $\epsilon t$, its behavior cannot be accurately captured using a single-scale analysis. MMS addresses this by introducing a hierarchy of independent time variables (e.g., $T_0$=t, $T_1$=ϵt, $T_2$=ϵ$^2$t, etc.), which allows the separation of dynamics occurring at different rates. This separation eliminates secular terms from the solution and ensures a uniformly valid approximation over extended times.

Instead of treating $x$ solely as a function of time $t$, MMS reformulates it as a function of multiple time scales $T_0$, $T_1$, $T_2$,…, each representing a distinct order of the perturbation parameter $\epsilon$. The solution is then expanded as a power series in $\epsilon$, enabling a systematic approximation of the system's behavior across different temporal dynamics,

$$x = x_0 + \epsilon x_1 + \epsilon^2 x_2 + \cdots \tag{2}$$

Using chain rule:-

$$\frac{d}{dt} = D_0 + \epsilon D_1 + \epsilon^2 D_2 \tag{3}$$

$$\frac{d^2}{dt^2} = {D_0}^2 + 2\epsilon D_0 D_1 + \epsilon^2({D_1}^2 + 2D_0 D_1) \tag{4}$$

Plugging in the values from equation (2), (3) and (4) in equation (1) while restricting to the order of O(ϵ), un-forced VPE transformed to following form,

$$({D_0}^2 + 2\epsilon D_0 D_1)(x_0 + \epsilon x_1) + \omega^2(x_0 + \epsilon x_1) = \epsilon(1 - (x_0 + \epsilon x_1)^2)(D_0 + \epsilon D_1)(x_0 + \epsilon x_1) \tag{5}$$

Thus, by using MMS, we transformed the given VPE from ODE to PDE (5) in terms of multiple scales. Order of magnitude analysis gives the following equations by equating terms of order *O(1)* and *O(ϵ)*,

$$O(1): \quad D_0^2 x_0 + \omega^2 x_0 = 0 \tag{6}$$

$$O(\epsilon): \quad D_0^2 x_1 + \omega^2 x_1 = -2D_0 D_1 x_0 + D_0 x_0 - x_0^2 D_0 x_0 \tag{7}$$

Equation (6) general solution can be written as,

$$x_0 = a(T_1, T_2)\cos(\omega T_0 + \beta(T_1 T_2)) \tag{8}$$

Re-writing above equation in exponential function as,

$$x_0 = A(T_1, T_2)e^{i\omega T_0} + c.c, \tag{9}$$

$$where\ A = \tfrac{1}{2} a e^{i\omega T_0},\ a = amplitude \text{ and } \beta = \text{phase shift.} \tag{10}$$

Expanding equation (7) by substituting (9), we obtain,

$$D_0^2 x_1 + \omega^2 x_1 = -(2\omega D_1 A - \omega A + \omega A^2\bar{A}) i e^{i\omega T_0} - i\omega A^3 e^{3i\omega T_0} + c.c \tag{11}$$

At this stage, the presence of a secular term—proportional to $e^{i\omega T_0}$—would lead to unbounded growth in the solution, thereby violating the assumption of a uniformly valid expansion. To suppress this non-physical behavior and preserve the validity of the perturbation series, the coefficient of the resonant term must be set to zero, as follows,

$$(2\omega D_1 A - \omega A + \omega A^2\bar{A}) = 0 \tag{12}$$

The resulting modulation equation governs the slow-time evolution of amplitude and phase. By applying the method of separation of variables and transitioning to polar coordinates, we derive explicit expressions describing how the amplitude and phase of the solution evolve with time.

$$a = 2[1 + (\tfrac{4}{a^2} - 1)e^{-\omega\epsilon t}]^{-0.5} \tag{13}$$

$$\beta\,(T_0, T_1) = 0 \tag{14}$$

If we examine the equation (13), when t→ ∞, amplitude will approach to 2, indicating the presence of stable limit cycle for small values of nonlinear damping coefficient $\epsilon$. Furthermore, equation (14) indicates that for time scale $T_0$ & $T_1$, there is no variation in phase shift. Transforming equation (8) using equation (13) & (14), while ignoring particular solution of $x_1$, we obtain,

$$x = 2[1 + (\frac{4}{a^2} - 1)e^{-\epsilon t}]^{-0.5}\cos(\omega T_0) \tag{15}$$

This solution clearly represents a limit cycle with a steady-state amplitude of 2. Fig 1 compares the analytical approximation obtained via the Method of Multiple Scales (equation 15) with the numerical integration of the original Van der Pol equation (1) using a 4$^{th}$ and 5$^{th}$ order Runge-Kutta scheme.

As the nonlinear damping coefficient $\epsilon$ increases,

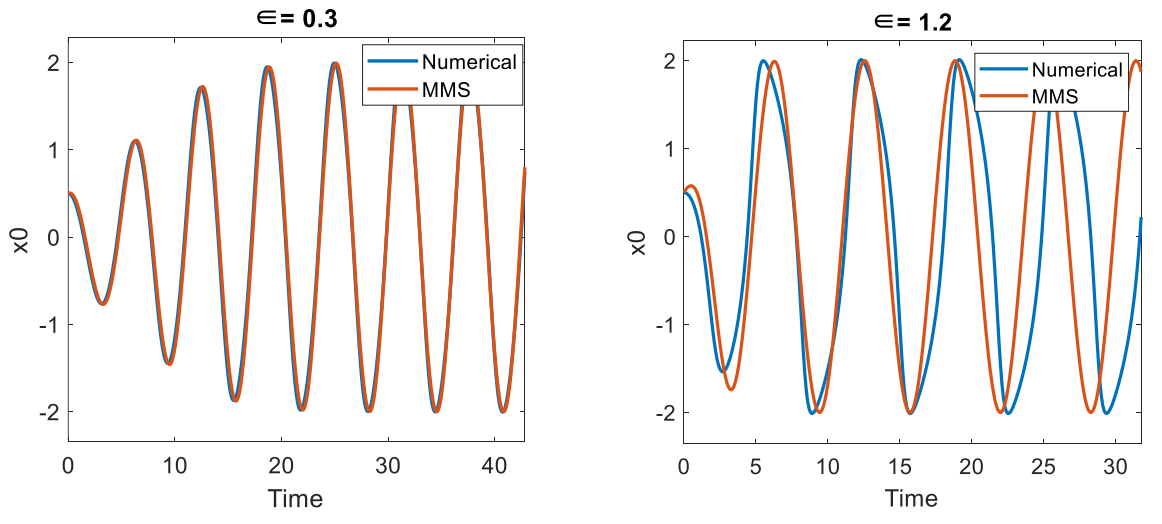

Fig. 1. Van der Pol equation analytical and numerical solution at $\epsilon = 0.3$ (Left) and $\epsilon = 1.2$ (Right)

discrepancies between the analytical and numerical solutions become more pronounced. This deviation arises from the fact that the analytical solution is truncated at *O(ϵ)*, neglecting higher-order effects. Nonetheless, over time, both approaches converge to a stable limit cycle with amplitude approximately equal to 2, consistent with the prediction from the amplitude expression in equation (13).

*B. Stability Analysis and Fixed Points (FP) Solution*

To delve deeper into stability analysis of the VPE, we can transform the understudy NL dynamical system into system of simultaneous ODEs by assuming $x = x_1$ , $\dot{x} = x_2$ , as follows,

$$\begin{pmatrix} \dot{x_1} \\ \dot{x_2} \end{pmatrix} = \begin{bmatrix} x_2 \\ -x_1 + \epsilon(1 - {x_1}^2)x_2 \end{bmatrix} \tag{16}$$

Fixed Points (FP) of the dynamical system obtained by equating to above mentioned ODEs (16) to zero, is, (0,0). In order to analyze the stability range of the system using FP analysis, we linearized the system to obtain corresponding Jacobian system at FP (0,0), as follows,

$$\begin{pmatrix} \dot{x_1} \\ \dot{x_2} \end{pmatrix} = \begin{bmatrix} 0 & 1 \\ -1 & \epsilon \end{bmatrix} \begin{pmatrix} x_1 \\ x_2 \end{pmatrix} \tag{17}$$

Which gives the characteristics polynomial as,

$$\rho^2 - \rho\epsilon + 1 = 0 \tag{18}$$

Corresponding eigen values given by,

$$\rho_{1,2} = \frac{\epsilon \pm \sqrt{(\epsilon^2 - 4)}}{2} \tag{19}$$

The stability characteristics of the fixed point can be determined by examining the eigenvalues of the Jacobian, derived from the linearized system (equations 18 and 19). For $\epsilon \geq 2$, both eigenvalues $\rho_{1,2}$ are real and positive, corresponding to an unstable node, as shown in Fig. 2(a). In the range $0 < \epsilon < 2$, the eigenvalues form a complex conjugate pair with positive real parts, indicating an unstable focus or spiral, illustrated in Fig. 2(b). When $\epsilon$=0, the eigenvalues are purely imaginary, signifying a non-hyperbolic center (Fig. 3(a)). As $\epsilon$ decreases into the negative regime, the fixed point becomes a stable spiral for $-2 < \epsilon < 0$, and transitions to a stable node when $\epsilon \leq -2$. These observations are made in the absence of external forcing and reflect how the system's local stability is governed by the sign and magnitude of the nonlinear damping coefficient. Notably, for $\epsilon < 0$, the damping is positive, leading to a globally stable response without the emergence of a limit cycle.

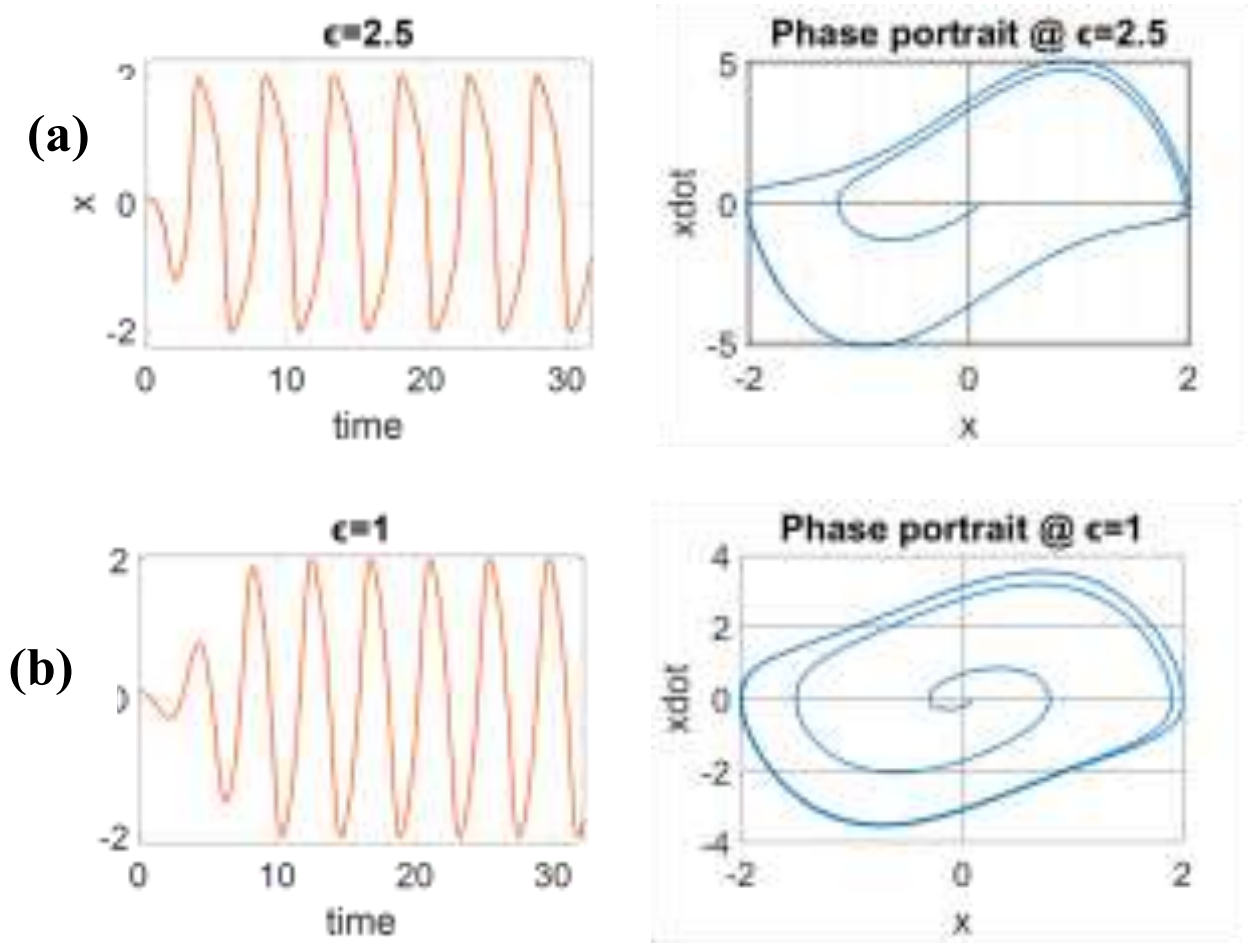

Fig. 2. Time series and phase portrait representations of the Van der Pol oscillator for different values of the nonlinear damping parameter $\epsilon$. (a) Unstable node observed at $\epsilon$=2.5. (b) Unstable focus observed at $\epsilon$=1.

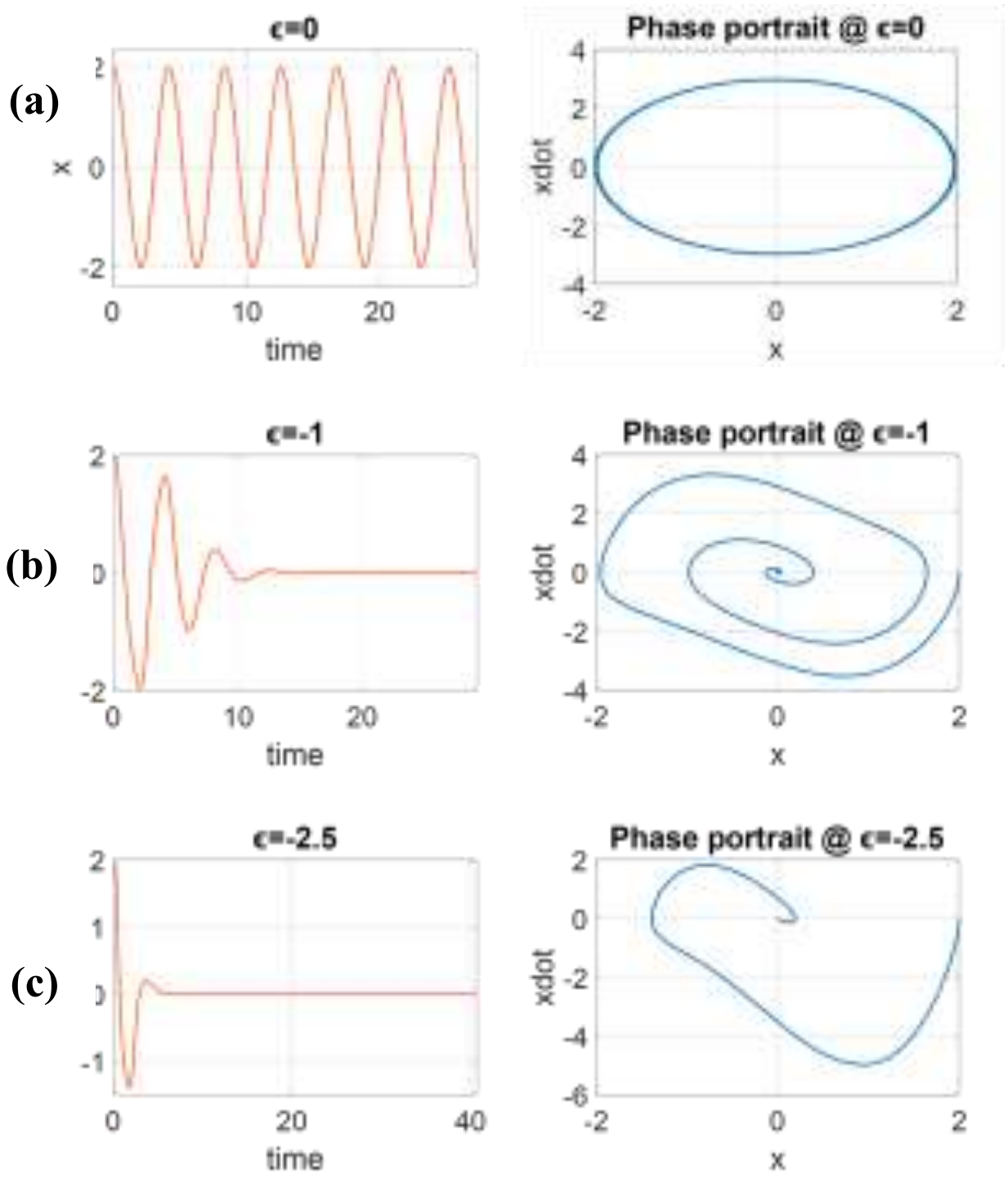

Fig. 3. Time series and phase portraits illustrating the behavior of the Van der Pol oscillator for negative and zero values of the damping parameter $\epsilon$.

(a) Non-hyperbolic center at $\epsilon$=0. (b) Stable Focus at $\epsilon$=-1 (c) Stable Node at $\epsilon$ =-2.5

For $\epsilon > 0$, numerical simulations confirm the emergence of a stable limit cycle with an amplitude of approximately 2, while the fixed point at the origin becomes unstable. This transition indicates the occurrence of a bifurcation at the equilibrium point (0,0) precisely when $\epsilon$ = 0. In the regime where $\epsilon < 0$, the fixed point is stable, and any limit cycle that exists is necessarily unstable. Conversely, for $\epsilon > 0$, the fixed point repels the trajectory, and the system settles into a stable limit cycle. This behavior highlights the classic property of a supercritical Hopf bifurcation: when the fixed point loses stability, a stable oscillatory solution emerges, and vice versa.

## III. QUALITATIVE DYNAMICS AND BIFURCATION

For $\epsilon > 0$, the system exhibits a periodic solution corresponding to a stable limit cycle, which acts as a global attractor. Regardless of whether the initial conditions lie inside or outside the limit cycle's trajectory, the system's response rapidly converges onto the periodic orbit. This behavior, illustrated in the phase portraits Fig. 4, confirms the robustness and stability of the limit cycle for a wide range of initial states.

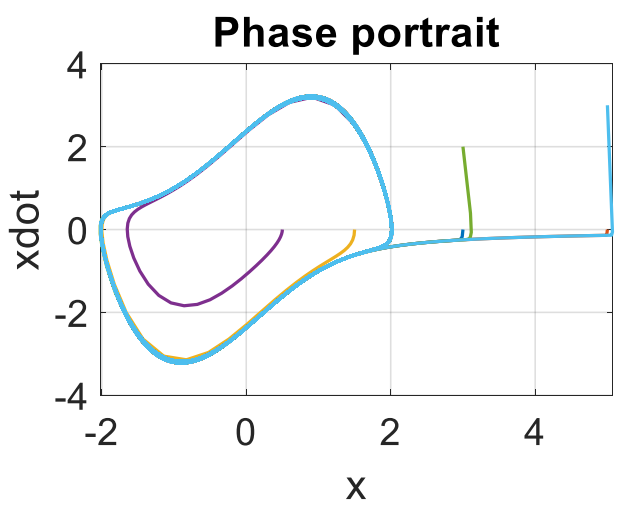

Fig. 4. Phase portraits of the Van der Pol oscillator for various initial conditions $(x, \dot{x})$ = (5,2), (5,0), (3,2), (3,0), (1.5,0), (0.5,0)

To characterize the nature of the bifurcation as the control parameter $\epsilon$ crosses the critical value $\epsilon_c$=0, we analyze the evolution of Floquet multipliers (FM) associated with the periodic orbit. A loss of stability is indicated when these multipliers move outside the unit circle in the complex plane. To compute them, two linearly independent solutions of the variational system (equations 20 & 21) were obtained by numerically integrating the system over the interval t∈[0,T], using the initial conditions $[1, 0]^T$ and $[0, 1]^T$. These solutions form the columns of the monodromy matrix Φ, whose eigenvalues provide the Floquet multipliers. The nature of the bifurcation is then inferred by examining how these multipliers behave as $\epsilon$ varies.

$$X_{10} = 2[1 + (\frac{4}{a^2} - 1)e^{-\epsilon t}]^{-0.5}\cos(\omega T_0)) \quad (20)$$

$$X_{20} = 2\omega[1 + (\frac{4}{a^2} - 1)e^{-\epsilon t}]^{-0.5}\cos(\omega T_0)) \quad (21)$$

$$\dot{y}_1 = y_2 \quad (22)$$

$$\dot{y}_2 = -y_1 + 2\epsilon y_1(a^2\omega\ \cos(\omega\, T_0)\sin(\omega T_0)) + y_2\epsilon(1 - a^2\cos^2(\omega\, T_0)) \quad (23)$$

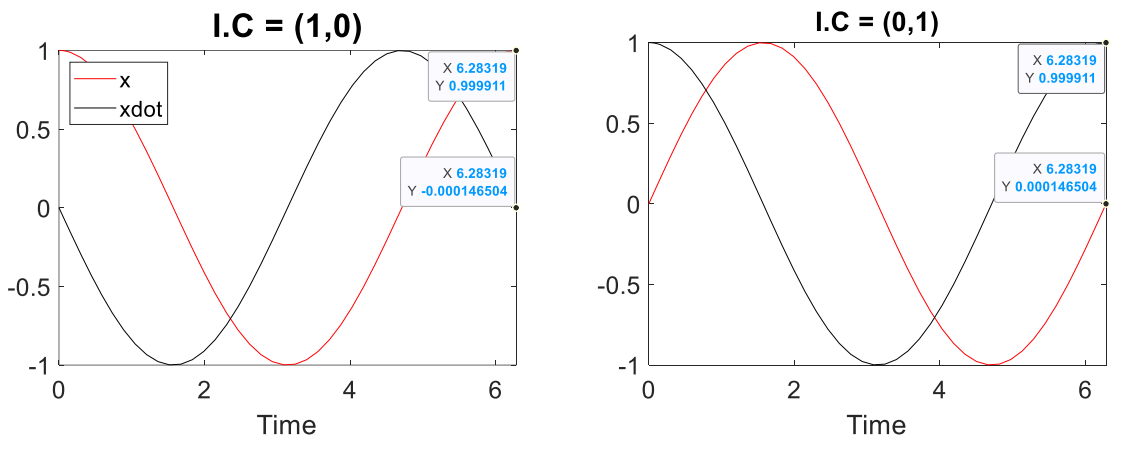

Fig. 5. Numerical solution of variational equations (22) and (23) over one time period **T = 2π/ω**

At $\epsilon$=0, the Floquet multipliers computed from the monodromy matrix (using equations 22 and 23) are both equal to 1, indicating that they lie precisely on the unit circle. This configuration signifies a non-hyperbolic periodic solution, which is marginally stable under linear perturbations.

The computed monodromy matrix Φ, obtained from the variational equations over one period, is given by:,

$$\Phi = \begin{bmatrix} 0.999 & 0.00014 \\ -0.00011 & 0.999 \end{bmatrix} \quad (24)$$

The corresponding Floquet multipliers obtained from the monodromy matrix are (0.999,0.999), both of which lie very close to unity. This indicates a near-marginally stable periodic solution at $\epsilon$=0. As the nonlinearity increases, the Floquet multipliers begin to deviate from the unit circle. For instance, at $\epsilon$=0.1 and $\epsilon$=0.5, the Floquet multipliers computed using equations (22) and (23) are given by:

$$FM_{\epsilon=0.1} = \begin{bmatrix} 1.3635 + 0.0094i \\ 1.3635 - 0.0094i \end{bmatrix} \quad (25)$$

$$FM_{\epsilon=0.5} = \begin{bmatrix} 4.4892 + 0.7022i \\ 4.4892 - 0.7022i \end{bmatrix} \quad (26)$$

As shown, the pair of Floquet multipliers move away from the unit circle in the complex plane when the control parameter $\epsilon$ exceeds the critical value $\epsilon_c$=0. This transition marks a secondary Hopf bifurcation, also known as a Neimark–Sacker bifurcation, which occurs in periodically forced systems. The nature of the resulting dynamics, whether periodic or quasi-periodic, depends on the relationship between the newly introduced frequency and the pre-existing fundamental frequency of the system. In this case, as shown in Fig 6, a new frequency component near 3 Hz emerges, while the original natural frequency remains at 1 Hz. This harmonic appears due to the cubic nonlinearity introduced in equation (11). Since the two frequencies are commensurate, the resulting post-bifurcation response remains periodic. However, with increasing nonlinearity (i.e., higher values of $\epsilon$), additional harmonics begin to manifest in the system's response, indicating increasingly complex dynamic behavior.

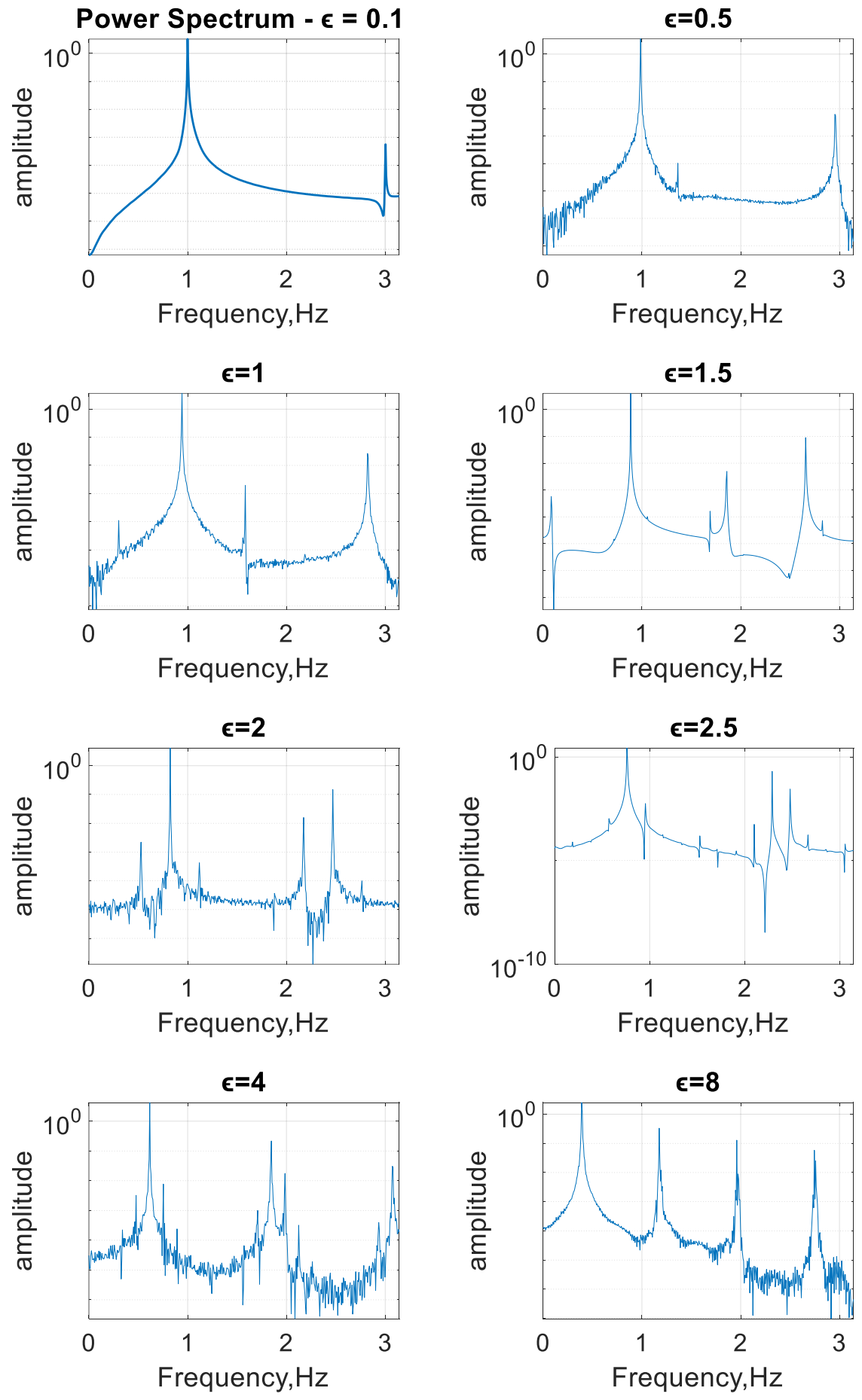

Fig. 6. Power spectrum of the system's response following bifurcation, showing dominant frequency components at $\omega_0$=1 and its harmonics. The emergence of additional peaks reflects the influence of nonlinearity in the post-bifurcation regime.

The harmonic structure exhibits a clear trend as the control parameter $\epsilon$ increases, as illustrated in Figures 6 and 7, respectively. The dominant frequency $f_1$, which corresponds to the natural frequency of the Van der Pol oscillator at small $\epsilon$, gradually decreases from 1 Hz at $\epsilon$=0.1 to approximately 0.389 Hz at $\epsilon$=8. With increasing nonlinearity, additional harmonic components appear in the spectrum, reflecting the growing complexity of the system's oscillatory behavior.

It is observed that while $f_1$ decreases with increasing $\epsilon$, the second and third harmonics ($f_2$ and $f_3$) approach one another up to $\epsilon$≈2 , after which they begin to diverge. A fourth harmonic becomes noticeable for $\epsilon$>4. At $\epsilon$=8, the spectrum also reveals the appearance of even harmonics, with a constant frequency spacing of $\Delta f \approx 0.78$ Hz. In the strongly nonlinear regime, the number and strength of these harmonics increase significantly, highlighting the rich spectral content generated by the system.

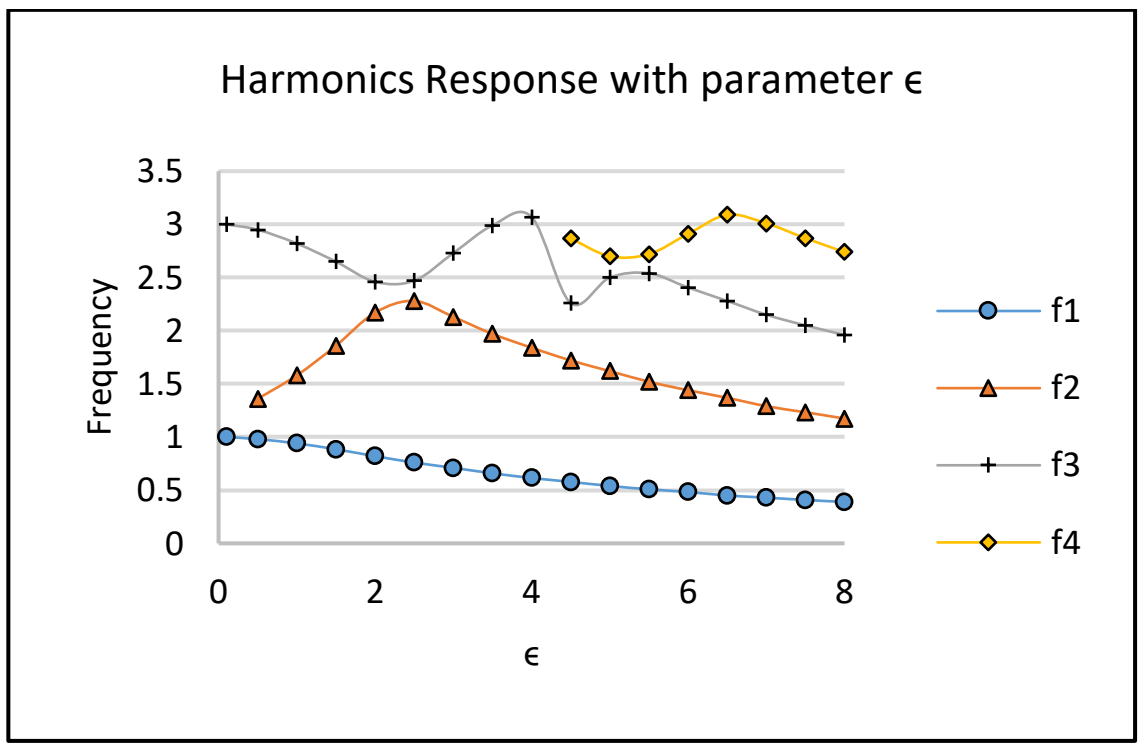

Fig. 7. Relationship between harmonic frequencies and the nonlinear control parameter $\epsilon$. The plot illustrates how $f_1$ decreases with increasing $\epsilon$, while higher-order harmonics emerge and evolve, reflecting the growing spectral complexity of the system.

## IV. Forced Van der Pol Equation

Using Forced Van der Pol equation with periodic forcing term $\boldsymbol{F}\cos(\omega t)$ with amplitude $\boldsymbol{F}$ and excitation frequency $\omega$, the dynamical system can be examined under the combined influence of the control parameters $\boldsymbol{F}$ and $\epsilon$. Numerical simulations show that the addition of forcing immediately induces a two-period quasi-periodic response in the system. System response is presented in Figure 8 for $\epsilon$=0.1, $\boldsymbol{F}$=0.5, $\omega_0$=1, and $\omega$=0.5.

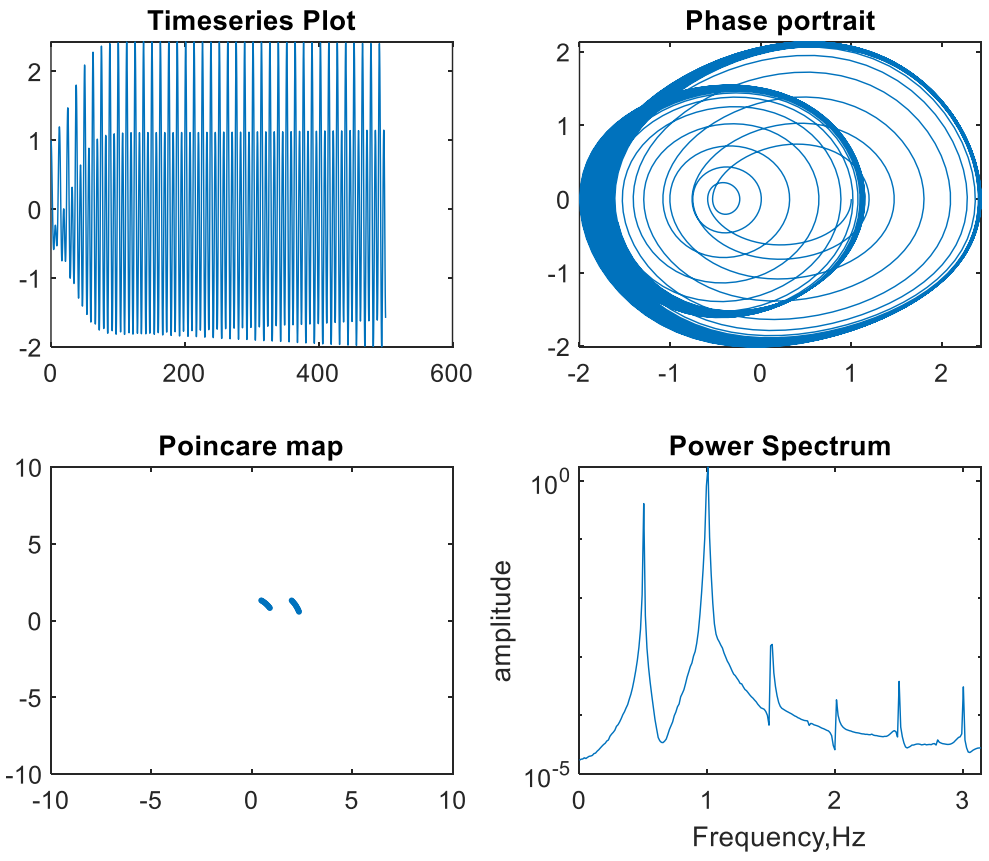

Fig. 8. Forced Van der Pol oscillator exhibiting a two-period quasi-periodic response for $\epsilon$=0.1, $\boldsymbol{F}$=0.5, $\omega_0$=1, and $\omega$=0.5.

For small forcing amplitudes, the system response contains both the natural frequency $\omega_0$=1 and the excitation frequency $\omega$=0.5. Even harmonics also appear in the spectrum. In the unforced Van der Pol oscillator, the emergence of such additional harmonics requires relatively high nonlinearity. In contrast, in the forced case, these harmonics arise even at low values of $\epsilon$ due to symmetry breaking. This behavior suggests that the external periodic forcing interacts with the system's intrinsic dynamics, lowering the nonlinearity threshold needed to induce symmetry breaking and enrich the spectral content. As shown earlier, in the absence of forcing, the system undergoes a Neimark bifurcation when $\epsilon > \epsilon_c = 0$, producing a periodic response that becomes quasi-periodic as $\epsilon$ increases. However, when an external forcing term is introduced, this transition changes in nature: instead of a periodic Neimark bifurcation, the system exhibits a two-period quasi-periodic bifurcation. To investigate this behavior in more detail, the system response was computed for $\epsilon$=0.5, $\omega_0$=1, and $\omega$ =0.5, while varying the forcing amplitude $\boldsymbol{F}$ over the values 0, 0.1, and 1, as illustrated in Fig. 9–11.

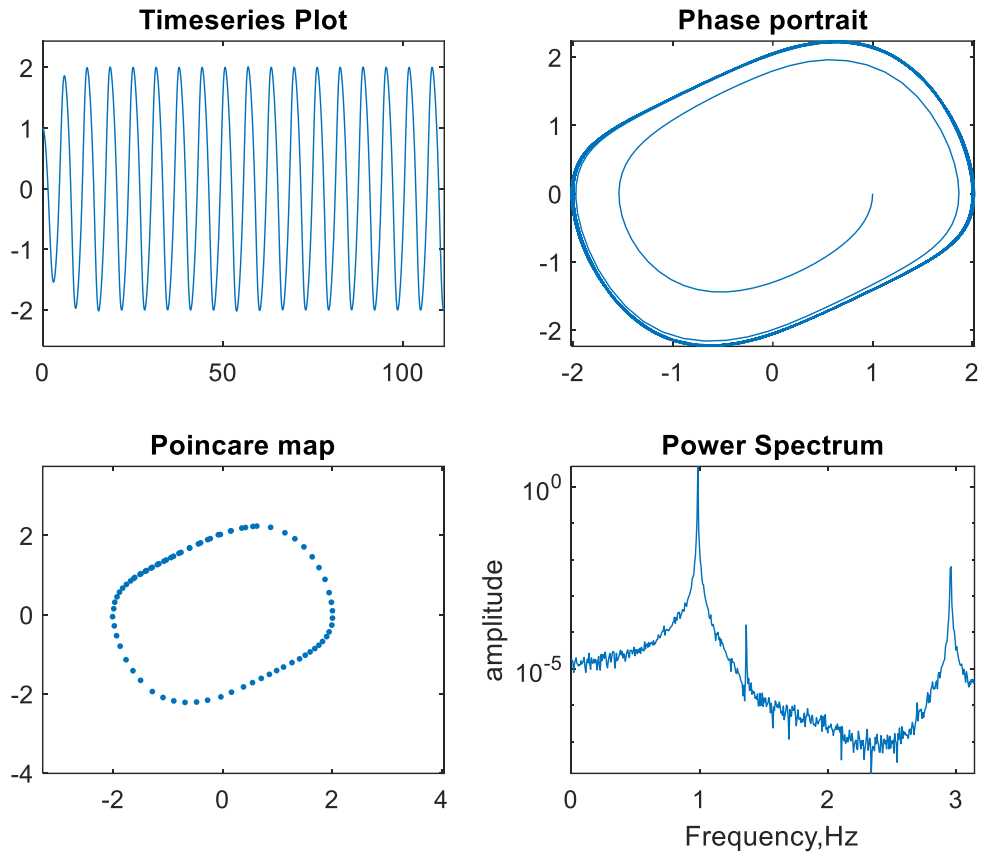

Fig. 9. Unforced Van der Pol oscillator response for $\boldsymbol{F}$=0, $\epsilon$=0.5

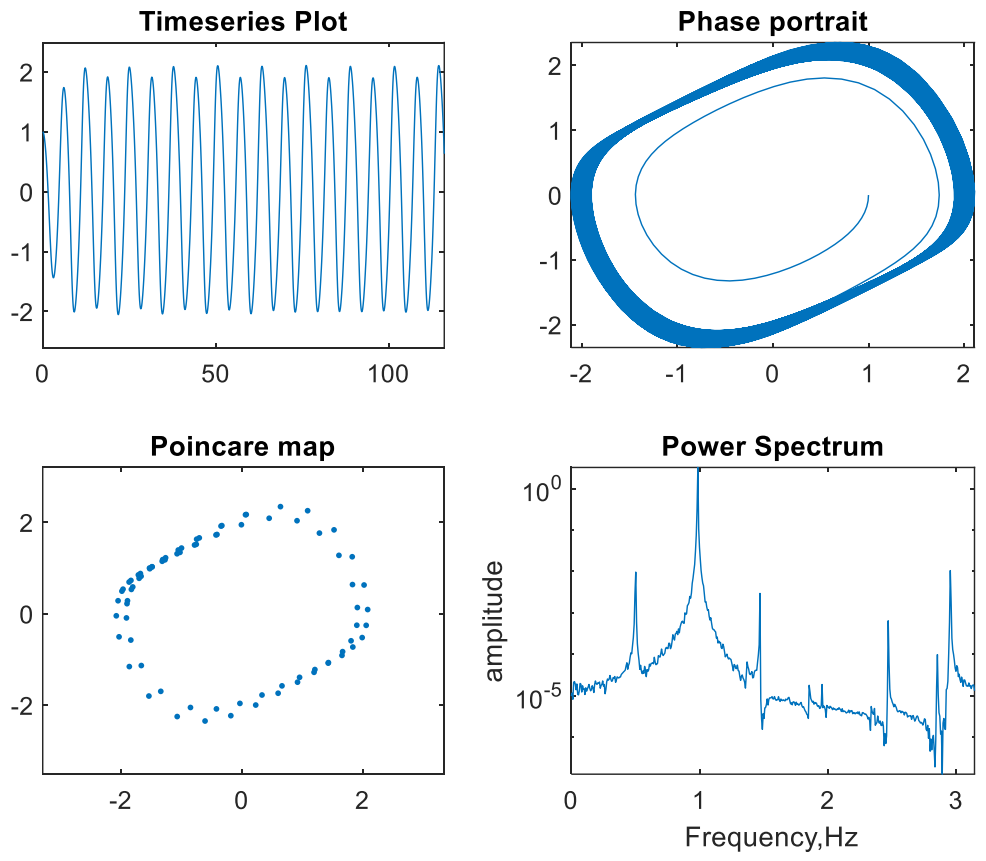

Fig. 10. Forced Van der Pol oscillator response for $\boldsymbol{F}$=0.1, $\epsilon$=0.5

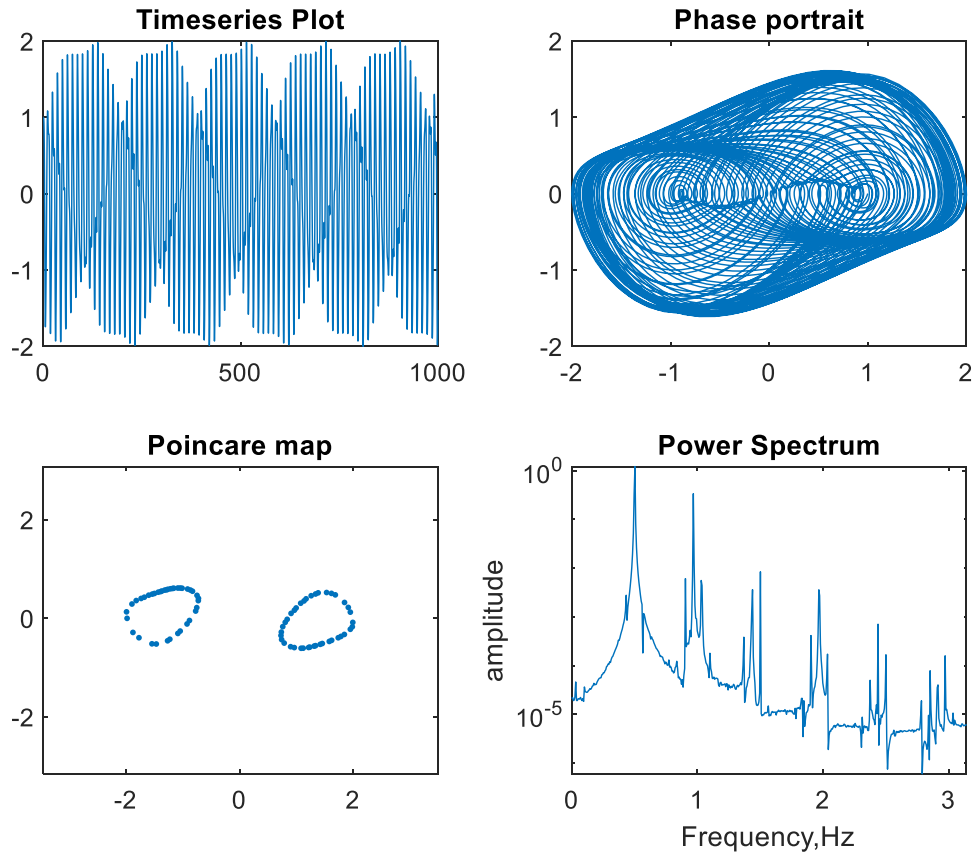

Fig. 11. Forced Van der Pol oscillator response for F=1, $\epsilon$=0.5

The Van der Pol oscillator exhibits clear co-dimension dependence at the critical values $\boldsymbol{F}_c$=0 and $\epsilon_c$=0. For $\boldsymbol{F}$=0 and $\epsilon_c$>0, the system undergoes a Neimark bifurcation, as identified earlier through Floquet multiplier analysis. When both $\boldsymbol{F}_c$>0 and $\epsilon_c$>0, the system instead transitions via a two-period quasi-periodic bifurcation, as illustrated in Fig. 10 and 11. For $\boldsymbol{F}_c$>0and $\epsilon_c$=0, the dynamics follow a period-doubling bifurcation route, as shown in Fig. 12. Increasing the forcing amplitude also promotes frequency locking with the excitation frequency. For example, at $\boldsymbol{F}$=0.1 (Fig. 10), the dominant frequency remains at 1 Hz ($\omega_0$=1), whereas at $\boldsymbol{F}$=1 (Fig. 11), the dominant frequency shifts to 0.5 Hz ($\omega$=0.5). Overall, three distinct bifurcation routes emerge for co-dimension parameters ($\boldsymbol{F}_c$,$\epsilon_c$) = (0,0):

1. Period-doubling ($\boldsymbol{F}_c$>0, $\epsilon_c$=0),
2. Neimark ($\boldsymbol{F}_c$=0 , $\epsilon_c$>0), and
3. Two-period quasi-periodic ($\boldsymbol{F}_c$>0, $\epsilon_c$>0),

each of which can ultimately lead to chaotic dynamics, as depicted in Fig. 13.

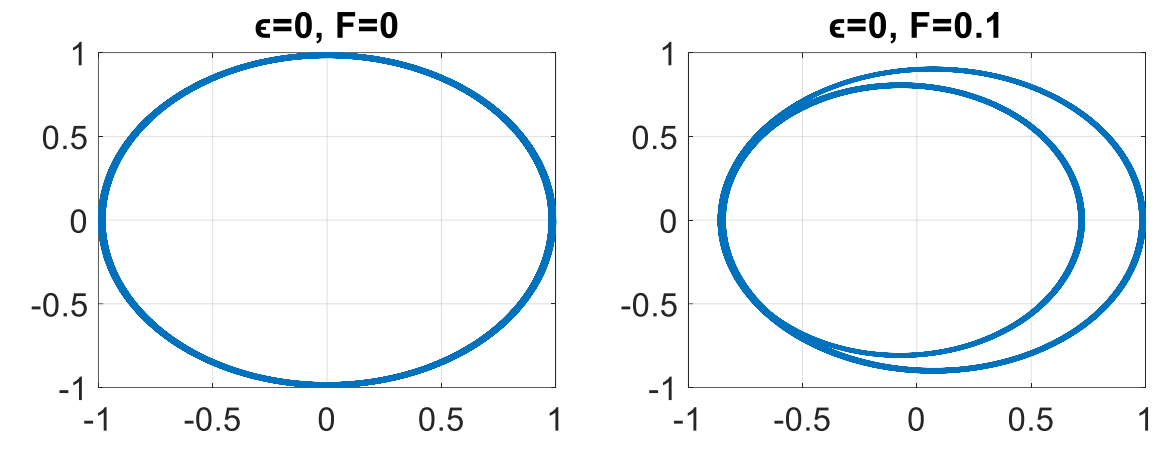

Fig. 12. Phase portrait of the Van der Pol oscillator showing a period-doubling bifurcation for forcing amplitude F > 0

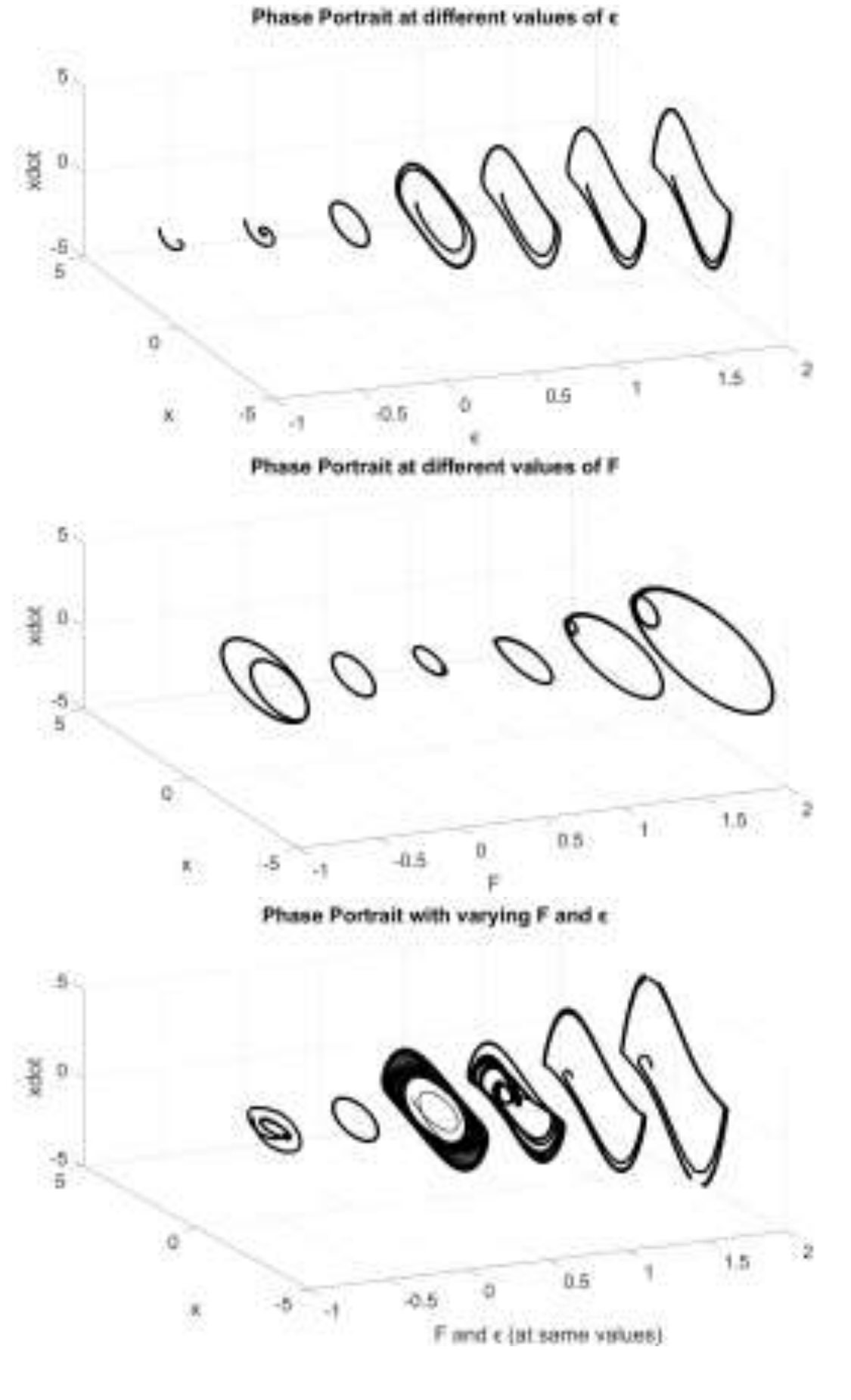

Fig. 13. Forced Van der Pol oscillator with ω = 1, $\omega_0$ = 0.5:
(a) Neimark bifurcation, (b) Period-doubling bifurcation, (c) Two-period qusai-periodic bifurcation.

To further investigate the routes to chaos and examine the synchronization of the system to external forcing, forced Van der Pol oscillator bifurcation analysis was analyzed. The system was solved via direct time integration, and the resulting bifurcation schematic comparison is presented in Fig. 14. Bifurcation diagrams were constructed by varying the forcing amplitude $\boldsymbol{F}$ from 0 to 2 while fixing the forcing frequency at ω =0.5. Numerical simulations were carried out for a range of damping parameter values, $\epsilon$ = [0.0, 0.1, 0.5, 1.0], to capture the system's dynamic transitions. After discarding transient dynamics, local maxima of the displacement x(t) were extracted to reveal the steady-state behavior. Notably, for $\epsilon \geq 0.5$, the system exhibits a clear sequence of dynamical transitions: beginning with simple periodic motion, undergoing period-doubling bifurcations, entering quasi-periodic and chaotic regimes, and then returning to a near-linear periodic response around $\boldsymbol{F} \approx 1\boldsymbol{F}$. Beyond this point, the system displayed a single local maximum per cycle with amplitude growing steadily with $\boldsymbol{F}$. This behavior is attributed to the external forcing becoming sufficiently strong to dominate the oscillator's intrinsic nonlinear dynamics, effectively entraining the system to the drive frequency and suppressing complex internal modes. In contrast, at lower $\boldsymbol{\epsilon}$ values, the system exhibits weaker nonlinear effects, and the transition to chaos is less pronounced.

In case of a periodically forced nonlinear oscillator, case in point the Forced Van der Pol dynamical system discussed above, there exist specific regions in the parameter space where the oscillator's response frequency becomes a rational multiple of the forcing frequency. These regions are known as Arnold tongues, a concept introduced by Vladimir Arnold in his seminal study of circle maps and mode-locking phenomena [12]. Each tongue shaped region corresponds to a distinct frequency-locking ratio p:q, where the oscillator completes p cycles for every q cycle of the external forcing, as shown in Fig 15. Within these regions, the system exhibits phase-locked periodic motion due to commensurate frequencies, characterized by a stable and repeating trajectory synchronized with the forcing. Outside these regions, the response is typically quasi-periodic, where two incommensurate frequencies interact, or can become chaotic under stronger forcing or high nonlinearity. The boundaries of Arnold tongues often mark the transition from orderly (periodic) to complex (quasi-periodic or chaotic) dynamics. To quantify the sensitivity of the system to initial conditions and detect the onset of chaos, the largest Lyapunov exponent (LLE) was analyzed. For a dynamic system with state vector $\mathbf{X}$(t) describing the dynamics of the system, the LLE explains the average divergence exponential rate of the adjoining particles trajectories in phase space. Mathematically, it is defined as:

$$\boldsymbol{\lambda} = \lim_{t\to\infty} \frac{1}{t}\, ln \left(\frac{\| \delta\boldsymbol{X}(t) \|}{\| \delta\boldsymbol{X}(0) \|}\right) \tag{28}$$

where $\delta\boldsymbol{X}(0)$ is an infinitesimal perturbation to the initial condition, and $\| \delta\boldsymbol{X}(t) \|$ is the distance between the original and perturbed trajectories at time $t$. In practice, this is approximated by periodically renormalizing the separation vector to avoid numerical overflow and summing the logarithmic divergence over time intervals, as shown in Fig. 16. A positive Lyapunov exponent ($\lambda$>0) indicates chaotic behavior, where nearby trajectories diverge exponentially. Conversely, $\lambda$=0 corresponds to quasi-periodic motion, and $\lambda$<0 signifies convergence to a periodic attractor.

<table>
<tr><th>ϵ</th><th>Bifurcation Diagram of Forced Van der Pol Oscillator</th><th>Phase Portrait</th></tr>
<tr><td>0.0</td><td>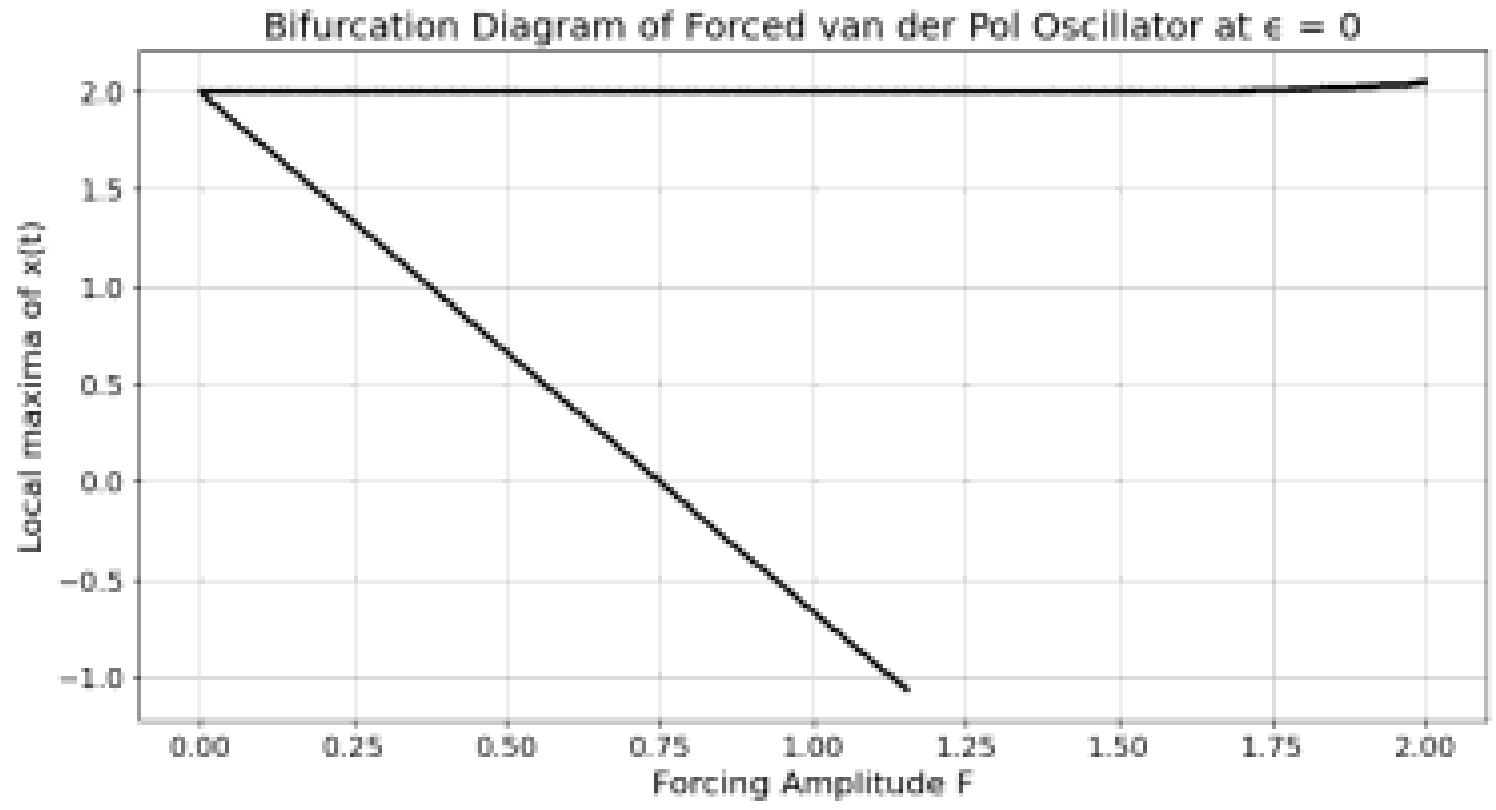
</td><td>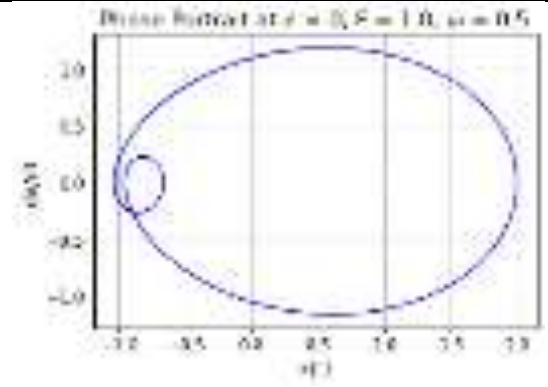<br>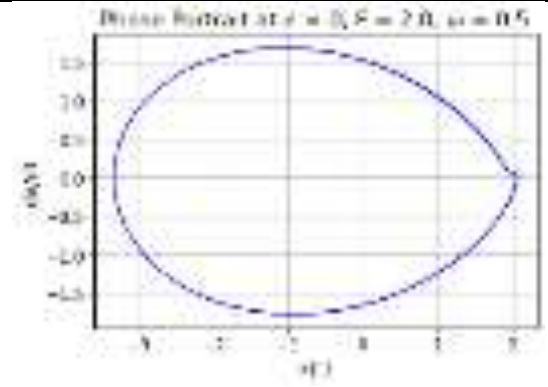</td></tr>
<tr><td>0.1</td><td>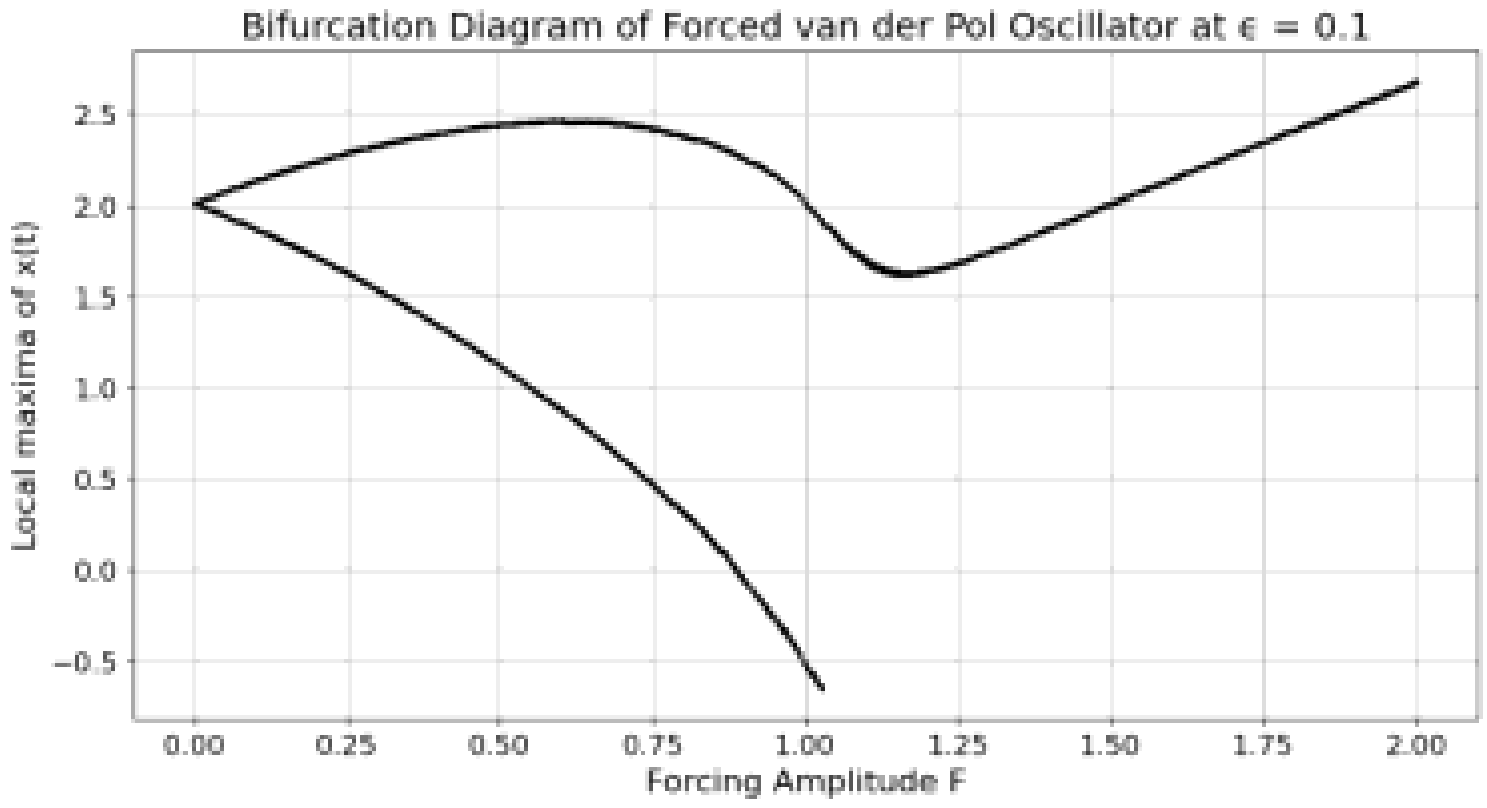
</td><td>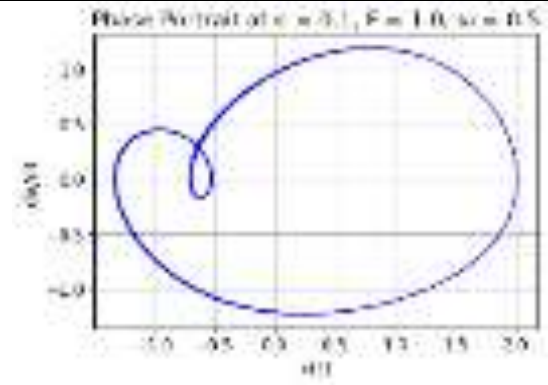<br>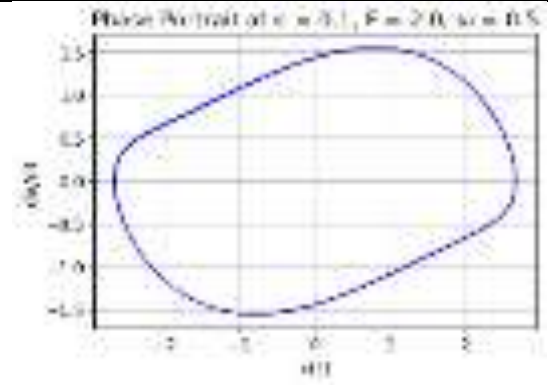</td></tr>
<tr><td>0.5</td><td>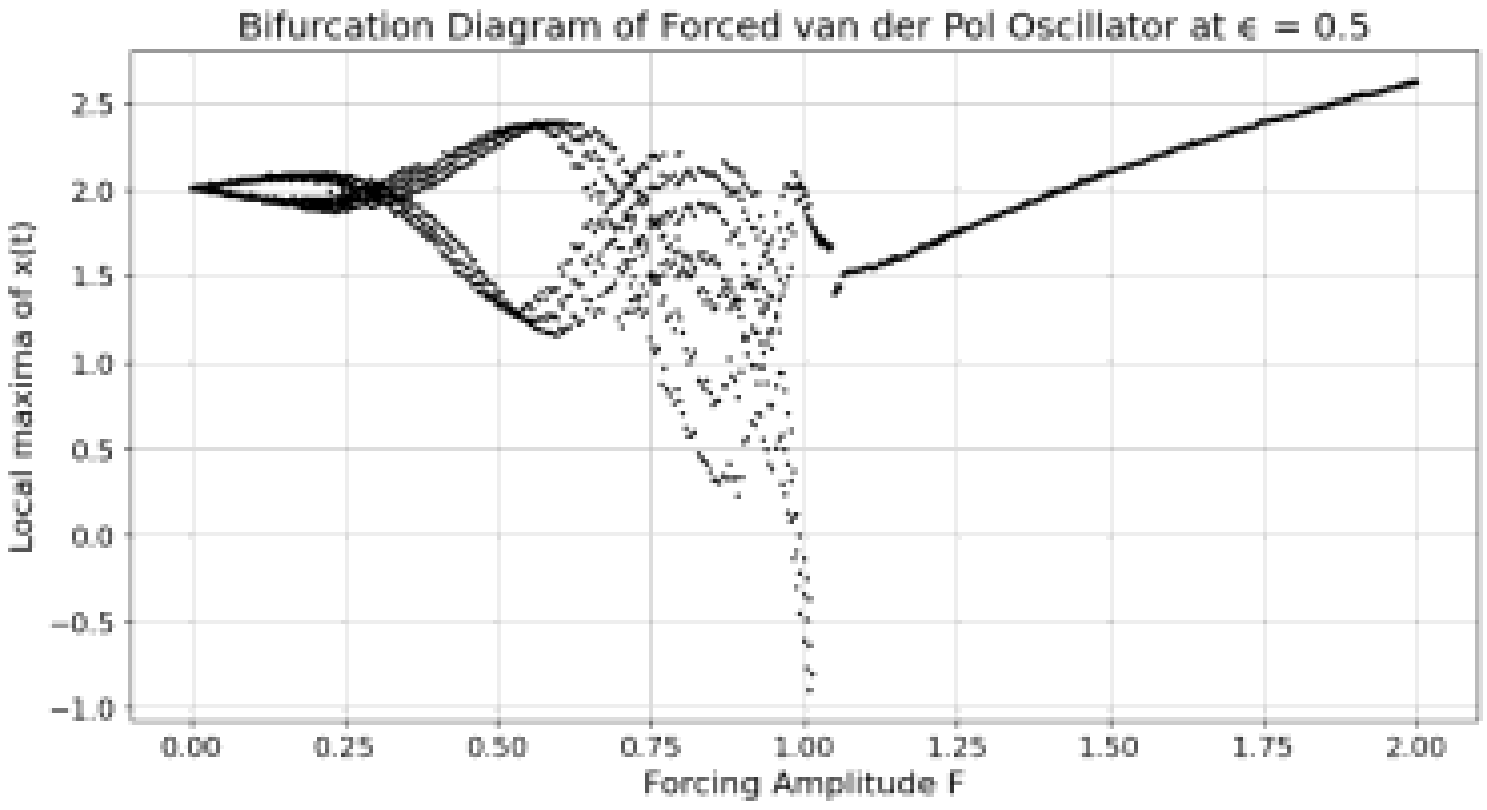
</td><td>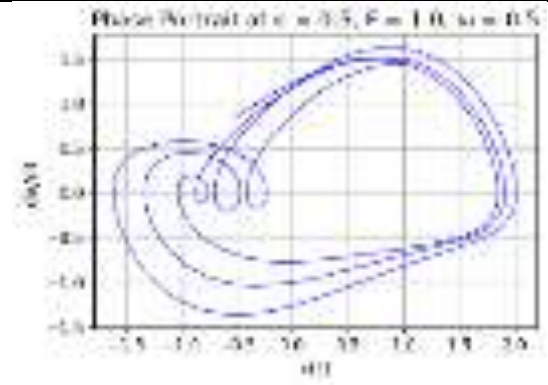<br>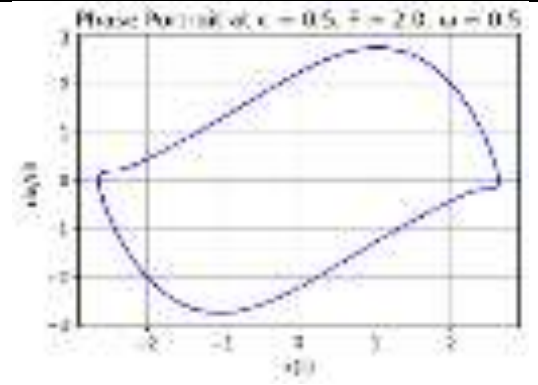</td></tr>
<tr><td>1.0</td><td>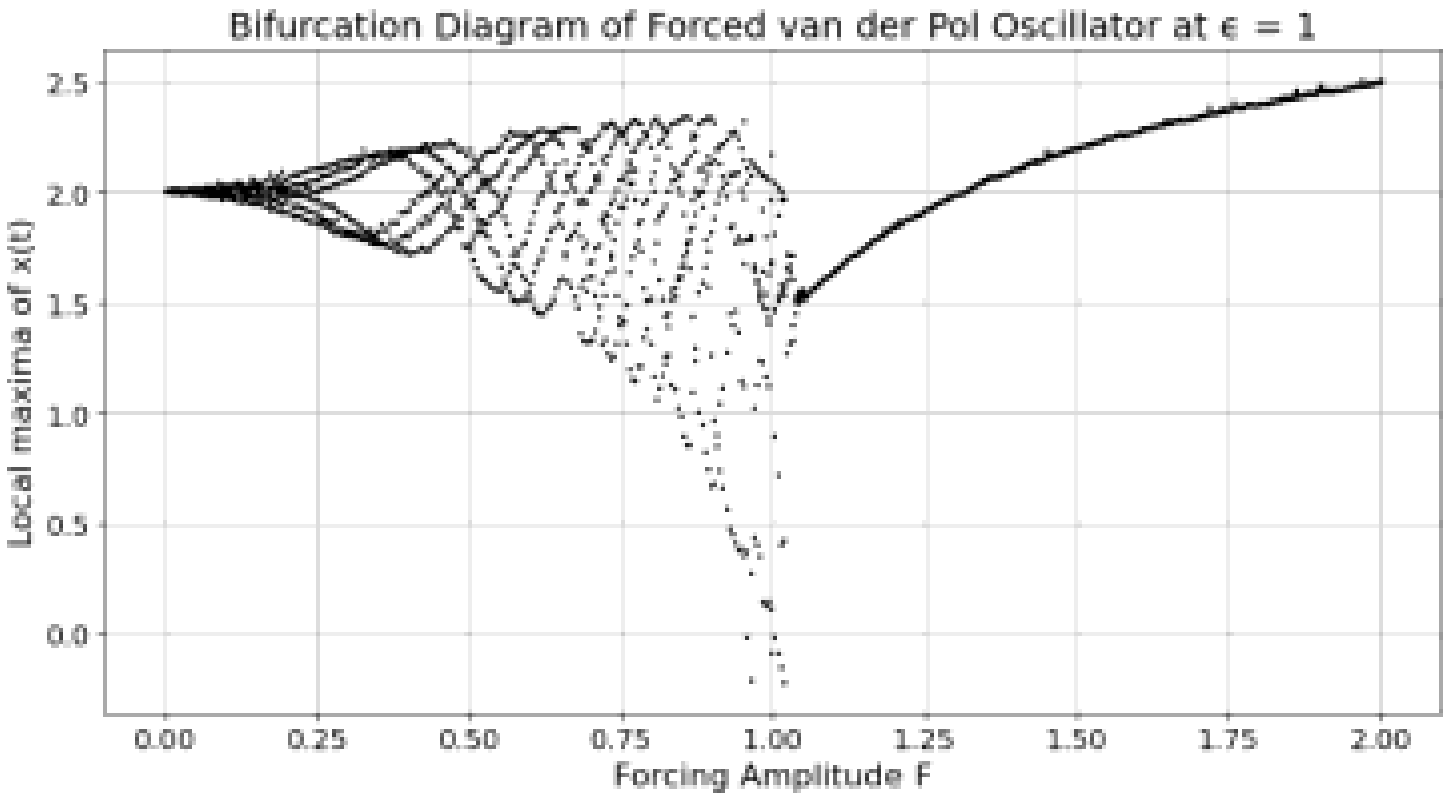
</td><td>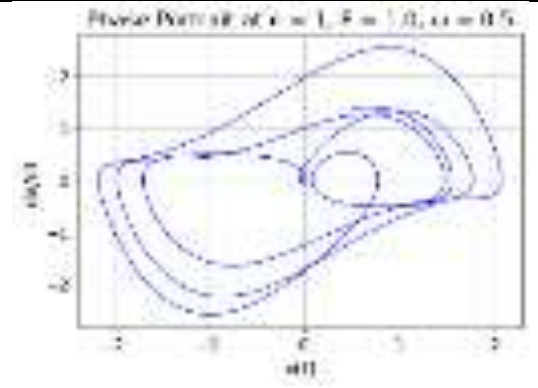<br>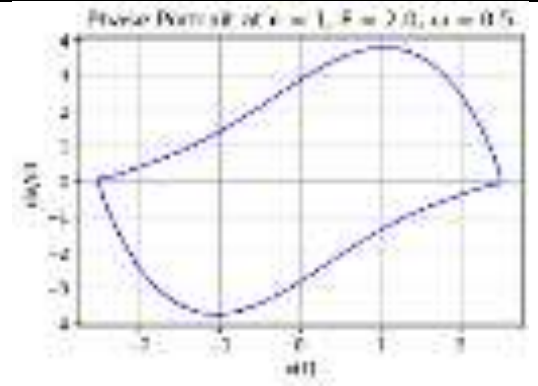</td></tr>
</table>

Fig. 14. Forced VPE maximum amplitude plot as a function of forcing function $\boldsymbol{F}$ @ $\omega = 0.5$, $\omega_0 = 1$ and varying degree of $\epsilon$. Each dot represents peaks per period. System depicts typical behavior of periodic oscillation to period doubling, quasi-periodic and chaotic behavior. Thereafter, as the forcing amplitude increased further from $\boldsymbol{F}\approx 1\boldsymbol{F}$ for high non-linear damping coefficient, the response dominated by the forcing component with periodic limit cycle oscillation.

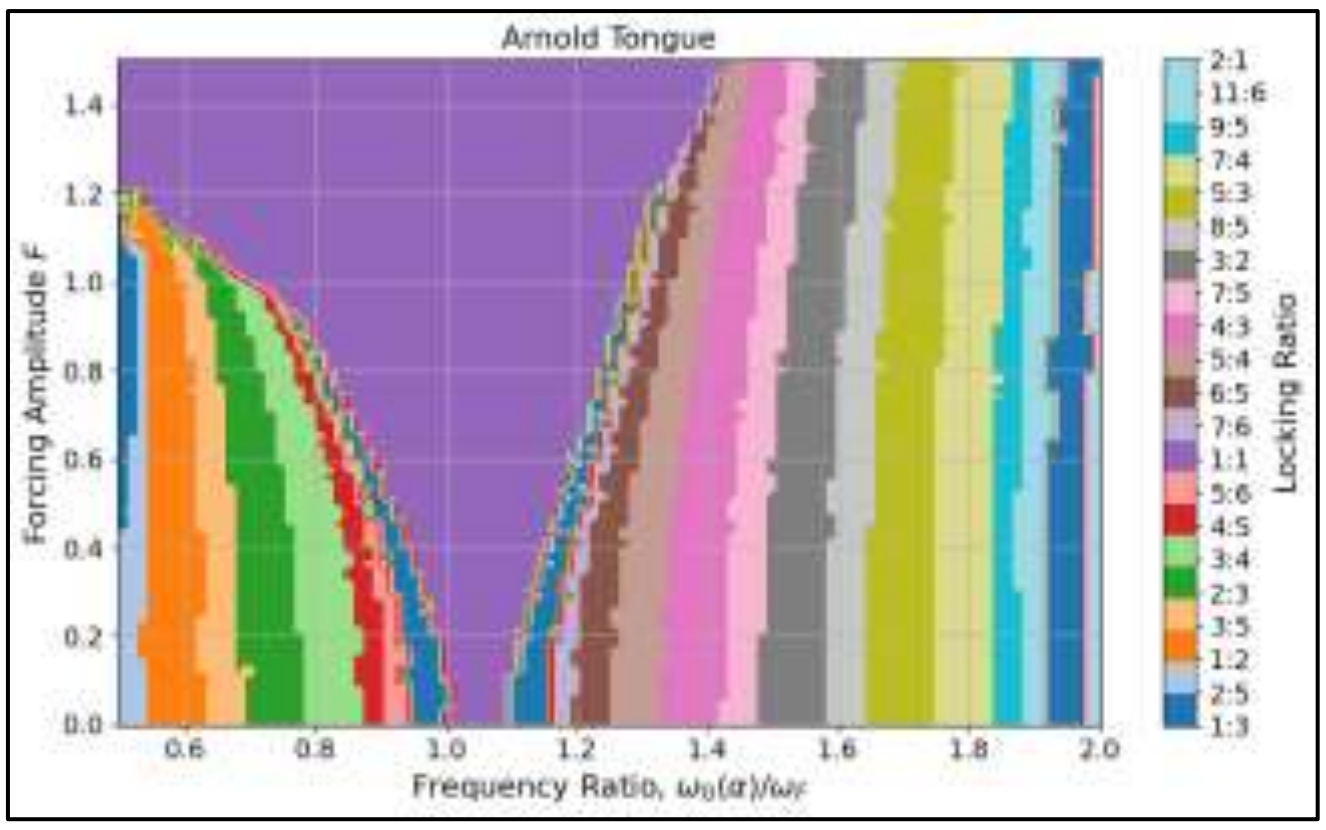

Fig. 15. Arnold tongue diagram for the forced Van der Pol oscillator showing frequency-locking regions (color-coded by locking ratio) as a function of forcing amplitude $\boldsymbol{F}$ and the ratio $\omega_0(\alpha)/\omega_F$. Each colored region corresponds to a different phase-locked region.

(a)

(b)

(c)

(d)

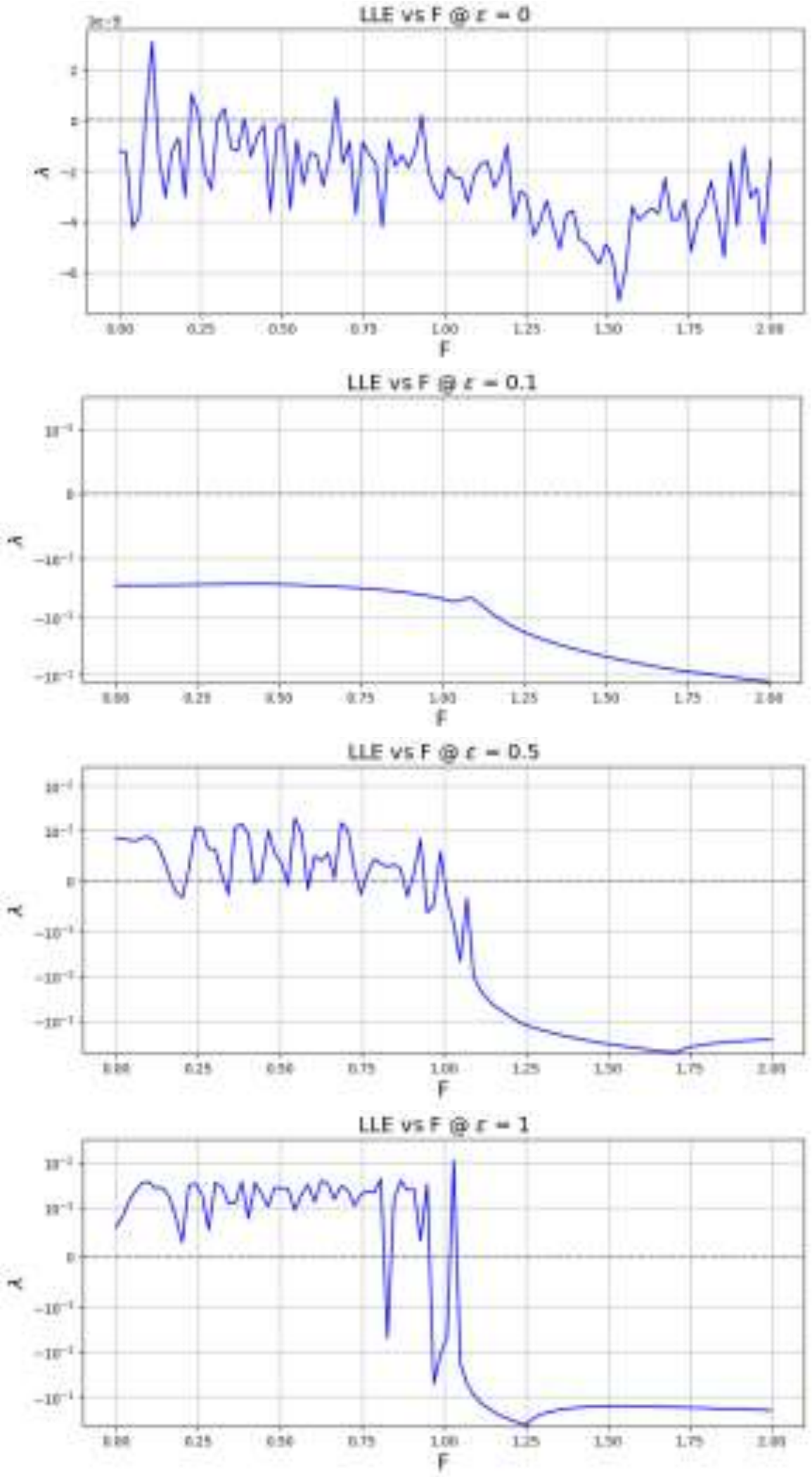

Fig. 16. Lyapunov exponent plot indicates onset of chaos as a function of forcing function $\boldsymbol{F}$ at increasing levels of $\boldsymbol{\epsilon}$. For $\boldsymbol{\epsilon} \geq 0.5$, LLE remains positive for $\boldsymbol{F} \leq 1$, indicating chaotic behavior as shown in fig 16(c) and (d).

## V. Van der Pol Oscillator application in Natural Phenomena/ dynamical system

The Van der Pol oscillator, which was initially conceived to describe triode oscillations in electrical circuits, has evolved into a fundamental model for understanding and simulating self-oscillatory systems across a remarkable range of scientific and engineering disciplines. Its significance stems from its ability to exhibit a stable limit cycle, representing sustained oscillations, and its capacity to display complex dynamics, including periodic to chaotic behaviors. This makes it a powerful tool for phenomenological modelling, capturing the essential nonlinear characteristics of many natural processes [13].

Beyond its electronic origins, the Van der Pol equation and its variants have found extensive use as basic models for oscillatory processes in physics, electronics, biology, neurology, sociology, and economics. For example, in heart dynamics, the Van der Pol oscillator is considered the simplest model for a natural pacemaker (sinoatrial node) because of its inherent limit cycle characteristic [14]. Van der Pol himself built electronic circuit models to study heart stability [15] [16], which later developed to study the effects of external driving, such as from a pacemaker, with the goal of stabilizing irregular rhythms (arrhythmias) [17]. More complex models of the cardiac conduction system utilize networks of coupled nonlinear oscillators, often based on or extending the Van der Pol framework. One approach models the heart's electrical system as a network of three nonlinear oscillators representing the SA node, the Atrioventricular node, and the His-Purkinje bundles, coupled by delayed connections and subjected to external stimuli [18] [19]. This multi-oscillator system generates a synthetic ECG signal, allowing for the study of normal and pathological rhythms through nonlinear analysis tools and experimental models using microcontrollers. Another method uses a modified Van der Pol oscillator model to replicate cardiac pulses from ECG signals, employing genetic algorithms for parameter optimization to match the morphological variability of individual heartbeats [20]. Piecewise analysis has also been applied to the Van der Pol equation to accurately replicate crossover behaviors observed in heart rhythm [21]. Biological rhythm modelling of cardiac dynamics in the context of heart physiology has typically been modeled as a stable limit cycle in various studies [19] [22] [23], analogous to the periodic motion observed at lower forcing amplitudes in our bifurcation diagrams. As parameters vary, the oscillator exhibits period-doubling, quasi-periodic, and chaotic behavior, which closely resembles the onset of cardiac arrhythmias observed in clinical data, such as those from the MIT-BIH Arrhythmia Database. The return to regular periodic oscillations at higher forcing amplitudes could be interpreted as analogous to the stabilization of heart rhythms through therapeutic interventions such as pacemakers. The numerical bifurcation patterns captured in our study not only highlight rich nonlinear dynamics but also provide a qualitative analogy to real-world biological phenomena.

Applications extend further, from modelling neurological defects in Parkinson's disease using delay differential equation representations of the Van der Pol model [17] to describing nonlinear aeroelastic instabilities such as Vortex-Induced Vibrations (VIV) and flutter in structures where the Van der Pol model amplitude-dependent aerodynamic damping [24]. In Earth sciences, the forced Van der Pol oscillator has been applied as a conceptual model for the dynamics of glacial cycles [25], and a hybrid Van der Pol-Duffing model describes the periodic nature of the solar magnetic field and sunspot numbers [26]. Its ability to model complex behavior also lends itself to novel approaches like physics-informed neural networks for forecasting extreme climatic events by embedding the oscillator's dynamics into deep learning frameworks [13].

## VI. Comparative Analysis

The Van der Pol equation has been studied using many approaches such as harmonic balance, averaging, continuation, and direct simulations. In this work, we applied the method of multiple scales (MMS) to approximate the amplitude and phase of oscillations and then validated stability through eigenvalue and Floquet analysis. The advantage of MMS is that it provides analytical relations that explain how nonlinear damping affects system response. However, this method is most reliable when nonlinearity is weak. Other methods offer different trade-offs. Harmonic balance and continuation techniques [27] can follow bifurcations across wider parameter ranges but may lose physical interpretability. Floquet analysis is powerful for identifying stability of periodic orbits, while direct numerical integration captures full nonlinear and chaotic behavior. Our study combines MMS with numerical results, which helps in both interpreting the dynamics and confirming their accuracy. Adding external forcing enriches the bifurcation structure. By tuning the nonlinear damping and forcing amplitude, one can observe Neimark bifurcations, period-doubling cascades, and two-frequency quasi-periodicity. Arnold tongue diagrams are often used to study frequency locking and synchronization under forcing. Lyapunov exponents provide a quantitative measure to confirm the transition to chaos. In short, our approach combines analytical clarity with numerical validation.

## VII. Limitation and future work

This study has some limitations. First, the perturbation approach is valid only for small values of the nonlinear parameter $\epsilon$, so it cannot fully describe strongly nonlinear cases. Second, we examined only a single oscillator; coupled or higher-dimensional systems could display richer dynamics. Third, the role of noise and uncertainty in parameters was not considered, though it is important in practical systems. Future work can address these issues. Global bifurcation tools and continuation methods can be used to map wider parameter ranges. Stochastic analysis can make the results more robust for real-world applications. In addition, data-driven methods such as symbolic regression or sparse identification (SINDy) could be applied to extract governing equations directly from measurements, for example from ECG signals, and then compared with the Van der Pol model.

## VIII. Conclusion

The Van der Pol oscillator provides a representative model for diverse physical systems that exhibit self-sustained oscillations. In this study, we highlighted the roles of bifurcation and limit cycles in shaping the system's qualitative behavior. Through exploration of parameter space and bifurcation diagrams, we showed that small variations in control parameters can induce significant shifts, including the onset of stable limit cycles. Perturbation methods, fixed-point stability analysis, and bifurcation theory proved effective for characterizing these nonlinear responses. More broadly, the analysis of dynamical systems often follows a structured pipeline: constructing bifurcation diagrams under multi-parameter variations, evaluating the largest Lyapunov exponent to detect transitions to chaos, and mapping frequency-locking regions through Arnold tongue diagrams. Together, these approaches form a robust framework for tracing routes to chaos and revealing complex nonlinear phenomena. Such a parametric analysis enables a deeper and more systematic understanding of the intrinsic dynamics of nonlinear systems.